\documentclass[12pt, a4paper]{article}
\pdfoutput=1

\usepackage{amsmath}
\usepackage{amsfonts}
\usepackage{amssymb}
\usepackage{adjustbox}
\usepackage{graphicx, rotating}
\usepackage{epstopdf}
\usepackage{epsfig}
\usepackage{latexsym}
\usepackage{color}
\usepackage{amsmath,bm,amssymb}
\usepackage{cite}
\usepackage{soul}
\usepackage{slashed}
\usepackage{float}
\usepackage[pagebackref=true]{hyperref}
\hypersetup{colorlinks, citecolor=bluscuro, linkcolor=bluscuro, urlcolor=bluscuro}
\definecolor{rossos}{cmyk}{0,1,1,0.55}
\definecolor{bluscuro}{rgb}{0.15, 0.2, .85}
\definecolor{bluchiaro}{cmyk}{1,.3,0.,0.1}
\definecolor{brown}{rgb}{0.6, 0.14, 0.14}

\newcommand{\LL}{\ell_{\Theta}}
\newcommand{\sgn}{\text{sgn}}

\newcommand{\be}{\begin{equation}}
\newcommand{\ee}{\end{equation}}
\newcommand{\bea}{\begin{eqnarray}}
\newcommand{\eea}{\end{eqnarray}}
\newcommand{\westcorner}{\mathbin{\rotatebox[origin=c]{45}{$\lrcorner$}}}

\newcommand{\eastcorner}{\mathbin{\rotatebox[origin=c]{225}{$\lrcorner$}}}
\newcommand{\southcorner}{\mathbin{\rotatebox[origin=c]{315}{$\lrcorner$}}}

\def\CS{\mathcal{S}}
\def\CC{\mathcal{C}}
\def\CW{\mathcal{W}}
\newcommand{\CO}{\mathcal{O}}
\def\CV{\mathcal{V}}

\begin{document}

\begin{titlepage}
\begin{flushright}
IFT-UAM/CSIC-19-156
\end{flushright}
\vspace{.3in}

\vspace{1cm}
\begin{center}
{\Large\bf\color{black} Entropic Locking Of Action Complexity At Cosmological Singularities}\\

\bigskip\color{black}
\vspace{1cm}{
{\large J.~L.~F. Barb\'on and  J.~Mart\'{\i}n-Garc\'{\i}a }
\vspace{0.3cm}
} \\[7mm]
{\it {Instituto de F\'{\i}sica Te\'orica,  IFT-UAM/CSIC}}\\
{\it {C/ Nicol\'as Cabrera 13, Universidad Aut\'onoma de Madrid, 28049 Madrid, Spain}}\\
{\it E-mail:} \href{mailto:jose.barbon@csic.es}{\nolinkurl{jose.barbon@csic.es}}, \href{mailto:javier.martingarcia1@gmail.com}{\nolinkurl{javier.martingarcia1@gmail.com}}\\
\end{center}
\bigskip

\vspace{.4cm}

\begin{abstract}
We study the relation between entropy and Action Complexity (AC) for various examples of cosmological singularities in General Relativity. The complexity is defined with respect to the causal domain of dependence of the singular set, and the entropy is evaluated on the boundary of the same causal domain. We find that, contrary to the situation for black hole singularities, the complexity growth near the singularity is controlled by the dynamics of the entropy $S$, with  a characteristic linear relation. This formula is found to apply  to singularities with vanishing entropy as well as  those with diverging entropy. In obtaining these results it is crucial to take into account the AC expansion counterterm, whose associated  length scale  must be chosen sufficiently large in order to ensure the expected monotonicity properties of the complexity.

\end{abstract}
\bigskip

\end{titlepage}


\section{Introduction}

\noindent

Quantum complexity has been proposed as a candidate quantity which would be capable of monitoring the interior of a black hole (cf. \cite{SusskindcomplexityandBHhorizons, StanfordShockWave, SusskindEntnotEnough} ). For a black hole of entropy $S$, its quantum complexity is expected to grow linearly over very long time scales, at least exponential in the entropy.   In the holographic map, this asymptotic linear growth of the quantum complexity corresponds to the growth of the interior geometry of the black hole, as measured by an extremal codimension-one surface (VC) or the on-shell action (AC) (cf. \cite{BrownSusskindAction, SusskindCAcorto}). 

The persistence of complexity growth well after the entropy of the black hole has stabilized is considered to be a fundamental property. More precisely, since the conjectured bound on complexity is non-perturbative in $1/S$, the bulk expansion parameter, the total AC/VC complexity accumulated by an eternal  black hole is infinite when computed in leading orders. This is  in correspondence to the infinite volume of an extremal hypersurface supported entirely in the interior, or the infinite on-shell action of the past domain of dependence of the singularity. 

It is interesting to check whether  entropy and complexity remain decoupled in more general situations where the entropy is not asymptotically constant. Given a singular spacetime of the type shown in Figure 1,   with a terminal singularity ${\cal S}^*$ and a horizon bounding the past domain of dependence $D^-({\cal S}^*)$, we can ask how the accumulated complexity compares with the entropy. In order to perform a sharp comparison, we must provide formal  definitions of both quantities adapted to this set up. More precisely, we consider the codimension-two sections ${\cal V}_u$, parametrized by a null coordinate $u$ which runs along the boundary of $D^- ({\cal S}^*)$, and a set of Wheeler-de Witt (WdW) patches, denoted ${\cal W}_u$, anchored on ${\cal V}_u$. Then, the natural comparison to perform is the volume of ${\cal V}_u$, measuring the entropy, versus the action of ${\cal W}_u$, measuring the AC complexity. 

 This AC prescription was dubbed `terminal AC' in \cite{terminals}, where some if its properties were studied.       
In the familiar case of a black hole with constant entropy, we may interpret the terminal AC  as a measure of the purely infrared contribution, including only those degrees of freedom which are actually involved in the holographic emergence of the black hole interior. In \cite{rabinovc, rabinoac, terminals}, this IR interpretation was generalized to situations with time-dependent entropy, corresponding to time-dependent UV/IR thresholds in AdS/CFT constructions. In these cases, the IR Hilbert space has a time-dependent dimension and the quasilocal AC gives a measure of the complexity accumulated in this time-dependent IR Hilbert space. 

In this paper, we regard the null boundary of $D^-({\cal S}^*)$ as carrying the `holographic data', essentially a Hilbert space of $u$-dependent dimension $\exp({\rm Vol}({\cal V}_u)/4G)$, and study the  complexity  assigned by the AC prescription, precisely in situations where ${\rm Vol}({\cal V}_u)$ varies strongly as $u\rightarrow u_*$ on approaching the singularity. Our results indicate that the rate of complexity growth is dominated by the finite-size effects in the Hilbert space, namely the rate of variation of the entropy, rather than the standard process of `entanglement weaving' which leads to linear complexity growth in black holes.

\begin{figure}[H]
\begin{center}
\includegraphics[width=6cm]{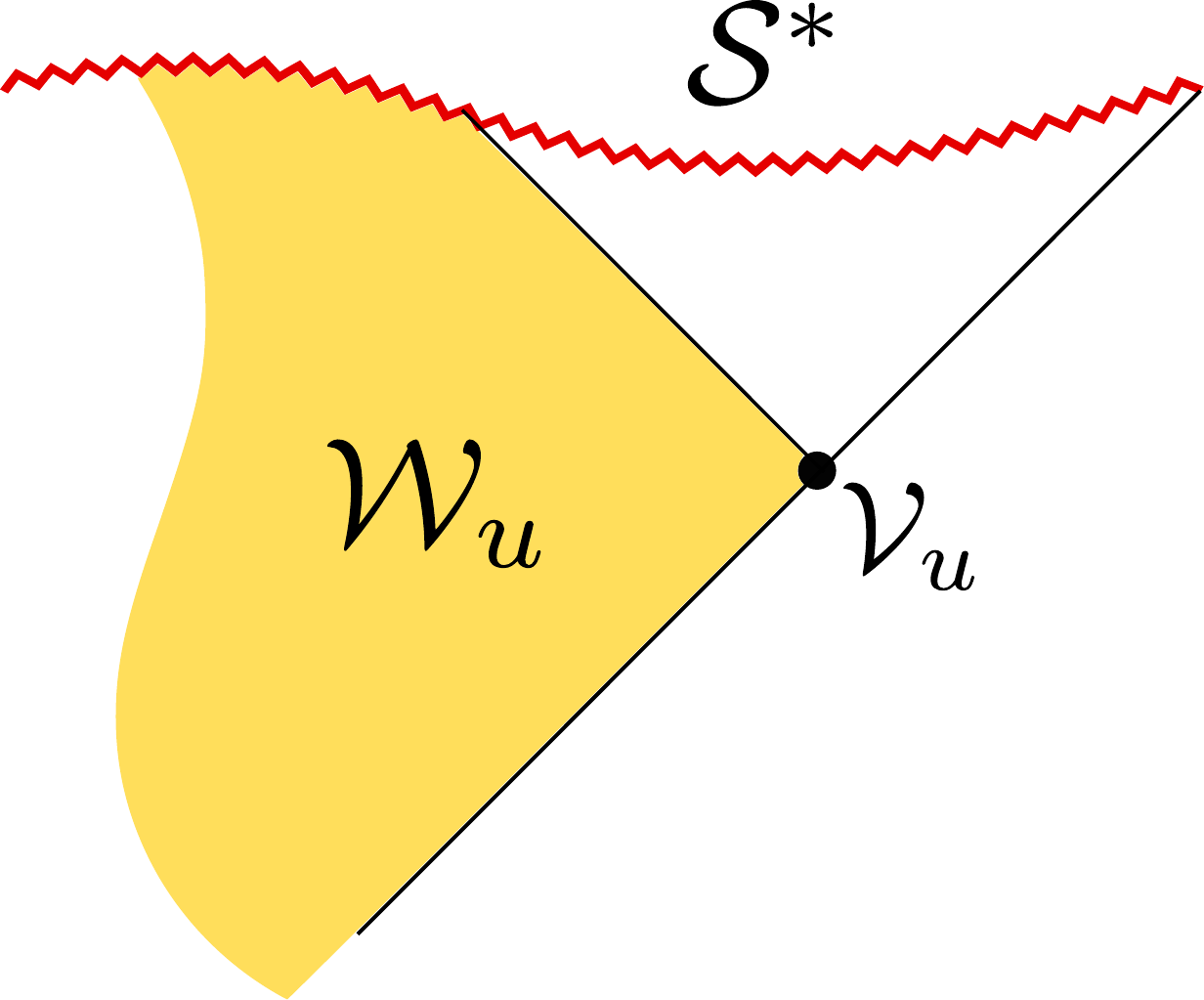} 
\caption{\emph{ The WdW patch $\CW_u$ (in yellow), parametrized by a null coordinate on the boundary of the singularity's past domain of dependence. Entropy is measured by the volume of ${\cal V}_u$ and complexity is measured by the action of ${\cal W}_u$.
This construction is modified in the obvious manner under time reversal.}}
\label{fig:terminalcomplexity}
\end{center}
\end{figure} 

This `entropic locking' of complexity is found in two qualitatively different situations  with nontrivial entropic behavior and still allowing an exact analytic treatment. One case corresponds to expanding bubbles of Coleman-de Luccia  type, engineered in concrete AdS/CFT scenarios.  Here both the terminal AC complexity and the entropy diverge at the singular locus and we are interested in the relative rates of divergence. The second example is a portion of  the Kasner spacetime, which is known to locally approximate any spacelike singularity  in GR. In this case the entropy vanishes and the terminal AC complexity approaches a constant.

We adopt a reparametrization-invariant prescription for the AC calculation, which requires the inclusion of the expansion counterterm, depending on a new length scale $\ell_\Theta$ (cf. \cite{Poisson}). A careful evaluation of this counterterm is crucial for our purposes, since we are precisely interested in situations with non-trivial null-expansion. 

Our main result is that as $u \rightarrow u_*$ and the singularity is approached in both classes of examples,  the  terminal AC growth is completely controlled by that of the entropy   $S$ through a law of the form
\begin{equation}
\label{law}
 \delta { \cal C^*} =  a\, \delta S+\dots  \,
\;,
\end{equation}
with $a$ a non-universal constant. Here, $S={\rm Vol} ({\cal V}_u)/4G$ and the dots stand for  $u$-independent contributions or subleading terms as $u\rightarrow u_*$.  More specifically, for the case of expanding bubbles, it is found that  the sign of the $a$ coefficient depends on $\ell_\Theta$, implying that a positive rate of complexity growth  actually requires  picking a sufficiently large value of this length scale, as measured in units of the AdS radius of curvature. In the Kasner case, the positivity of the coefficient $a$ is guaranteed by the weaker condition $\LL \gg \ell_{\text{Planck}}$ yielding a decreasing complexity as the dimensionality of the  effective Hilbert space is reduced.

This paper is organized as follows. In section 2 we review the terminal AC prescription, including the counterterm which ensures reparametrization invariance on the boundary of WdW patches. In section 3 we study the terminal complexity of singularities inside exact solutions of expanding bubbles with divergent entropy. In section 4 we compare the local behavior of AC complexity and entropy for a small patch of a generic spacelike singularity in GR, approximated as a Kasner metric. We end in section 5 with the conclusions. 

The computation of  terminal AC, with its implicit restriction to the causal past of the singularity, is technically  non trivial even for the standard case of AdS black holes. For this reason we have added an explicit discussion of these technicalities in an appendix.

\section{Terminal AC}
\label{sec:action}
\noindent

Given a terminal GR singularity ${\cal S}^*$ and a family of WdW patches ${\cal W}_u$ restricted to its past domain of dependence, we
define the quasilocal AC complexity as the on-shell action of the given WdW patches, supplemented by a countertem  $I_\Theta$ which restores reparametrization invariance: 
\begin{equation}\label{def}
{\cal C^*}_u \propto I[{\cal W}_u] + I_\Theta [{\cal W}_u]\;,
\end{equation}
where $I[{\cal W}_u]$ stands for the canonical gravitational action with a definite prescription for codimension-one and codimension-two boundary terms (cf. \cite{Poisson}). In particular 

\begin{eqnarray}
\label{masterformula}
16 \pi G I[\CW] &=& \int\limits_{ \CW} \text{d}^{d+1}x \,\sqrt{-g}\,(R-2\Lambda) \\ \nonumber &+& 2 \sum\limits_{T_i} \int\limits_{ T_i} \text{d}^dx \,K   + 2 \sum\limits_{S_i} \sgn(S_i)\int\limits_{ S_i} \text{d}^d x \,K  \\ &-&  2 \sum\limits_{N_i} \sgn(N_i)\int\limits_{ N_i} \text{d}^dx \,\text{d}\lambda \,\hspace{0.1cm}\kappa  +2 \sum\limits_{j_i} \sgn(j_i)\oint\limits \text{d}^{d-1}x \,\alpha_{j_i} \, , \nonumber
\end{eqnarray}
where in this expression

\begin{itemize}
\item $T_i$, $S_i$ and $N_i$ are respectively the timelike, spacelike and null boundaries of the WdW patch and $K$ are the traces of the corresponding extrinsic curvatures for the first two cases. For the null boundaries, $\lambda$ represents an arbitrary parameter on null generators of $N_i$, with $\kappa$ the surface gravity associated to $N_i$ in this parametrization. The sign for the null and spacelike boundary contributions are defined depending on the relative location of the boundary to the WdW patch as follows $$ \text{sgn} \left( \raisebox{0pt}{\includegraphics[height=0.26cm]{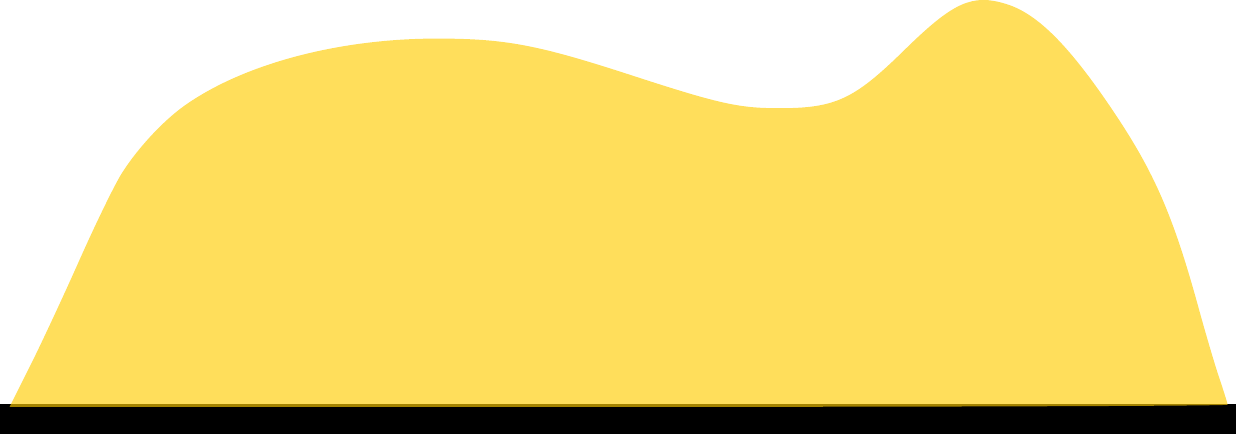}} \right)=\text{sgn} \left( \raisebox{-5pt}{\includegraphics[height=0.5cm]{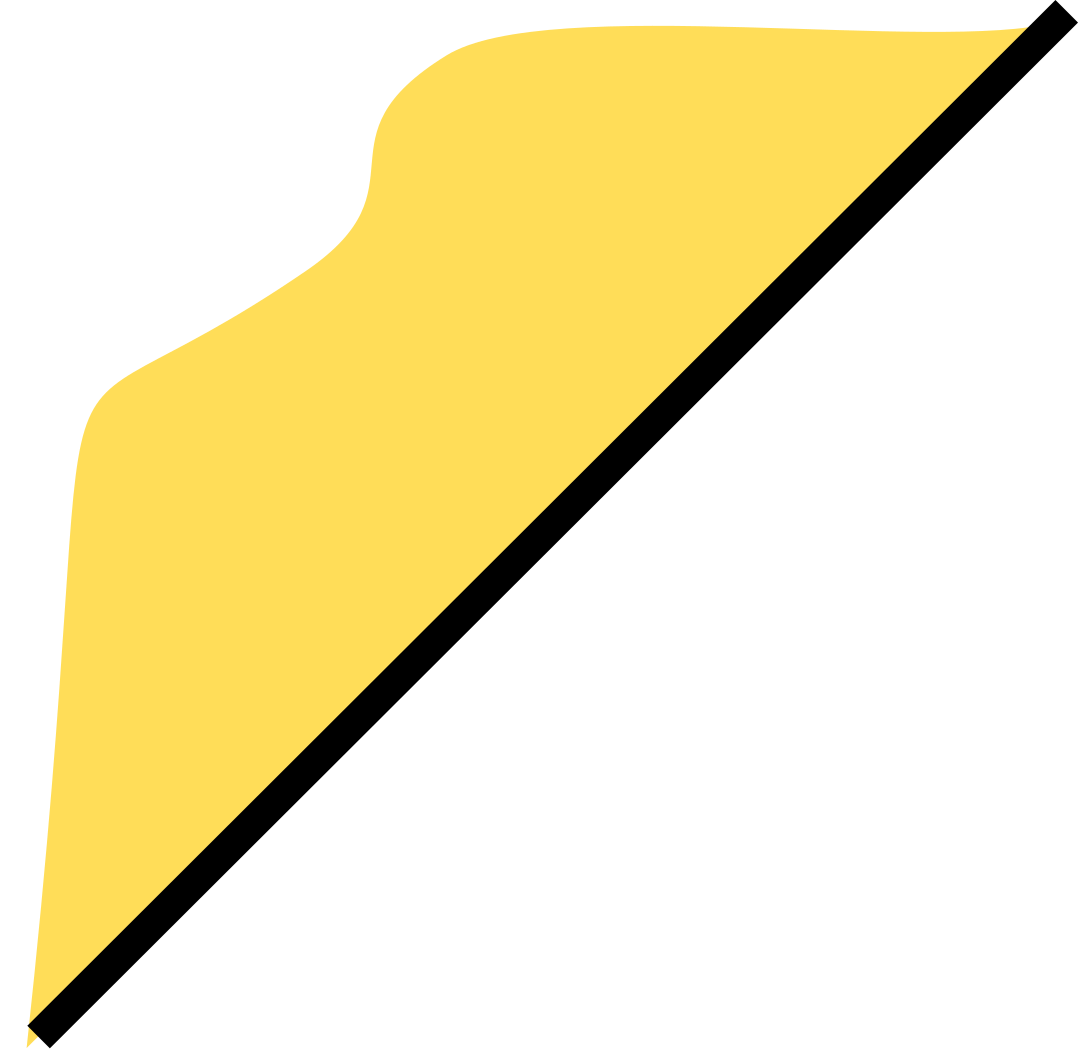}} \right)=\text{sgn} \left( \raisebox{-5pt}{\includegraphics[height=0.5cm]{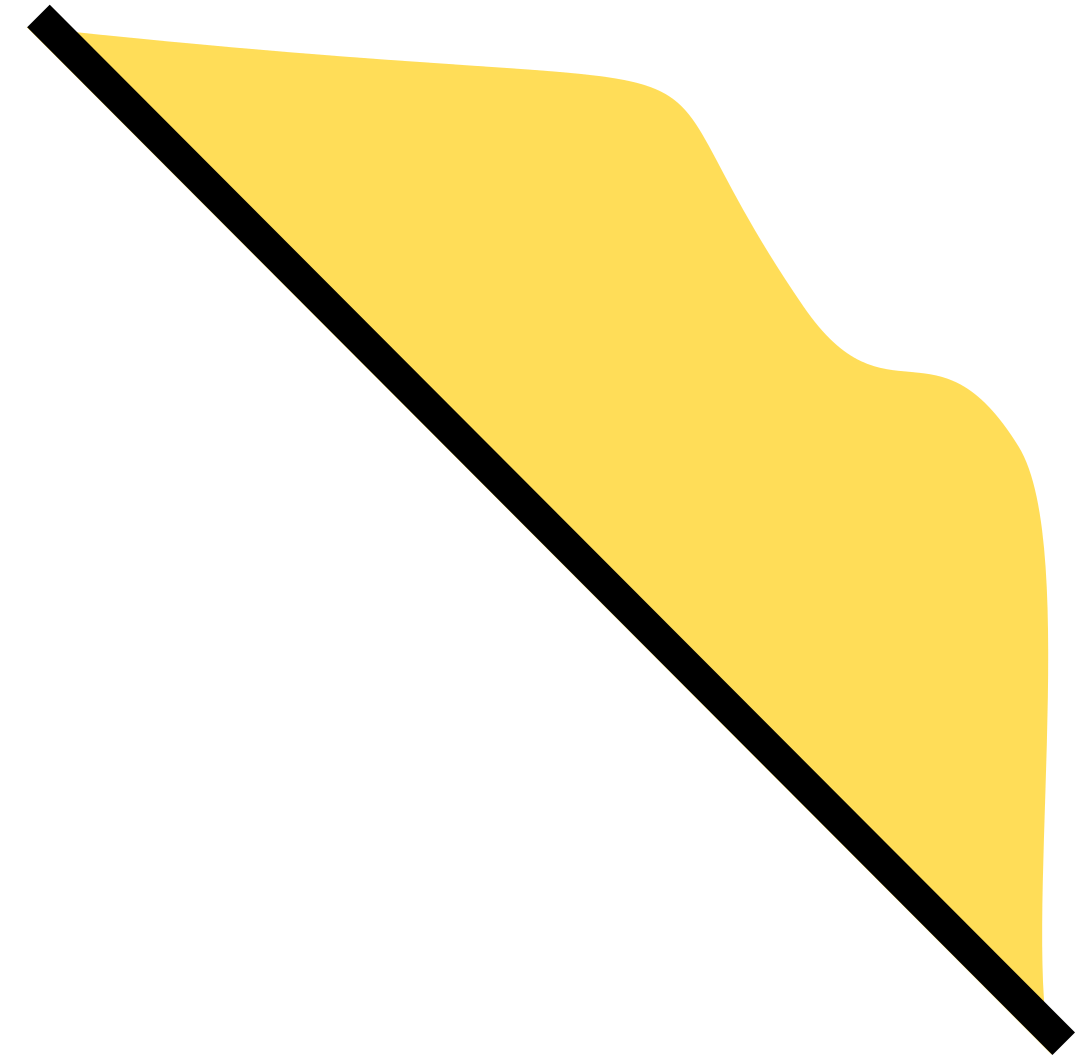}} \right)=-\text{sgn} \left( \raisebox{-2pt}{\includegraphics[height=0.26cm]{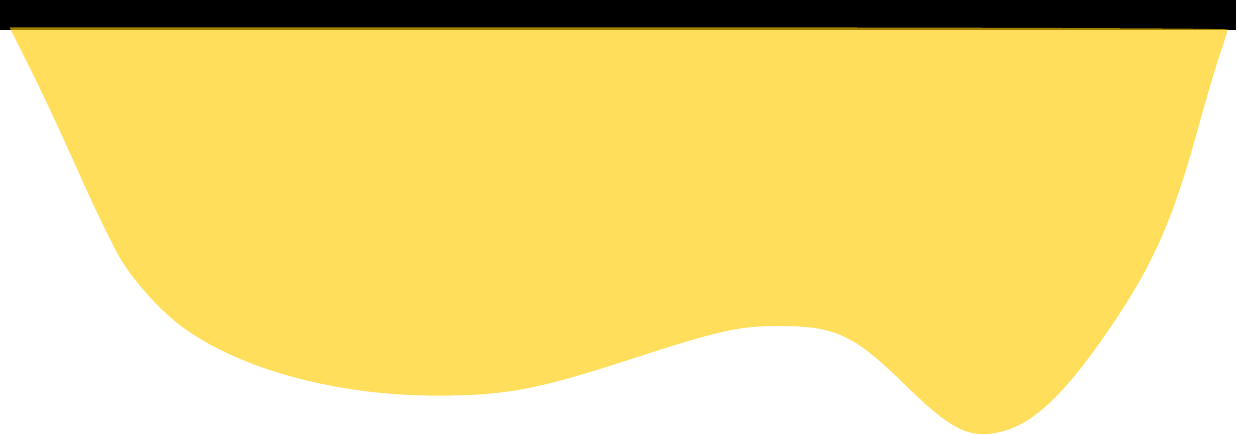}} \right)=-\text{sgn} \left( \raisebox{-5pt}{\includegraphics[height=0.5cm]{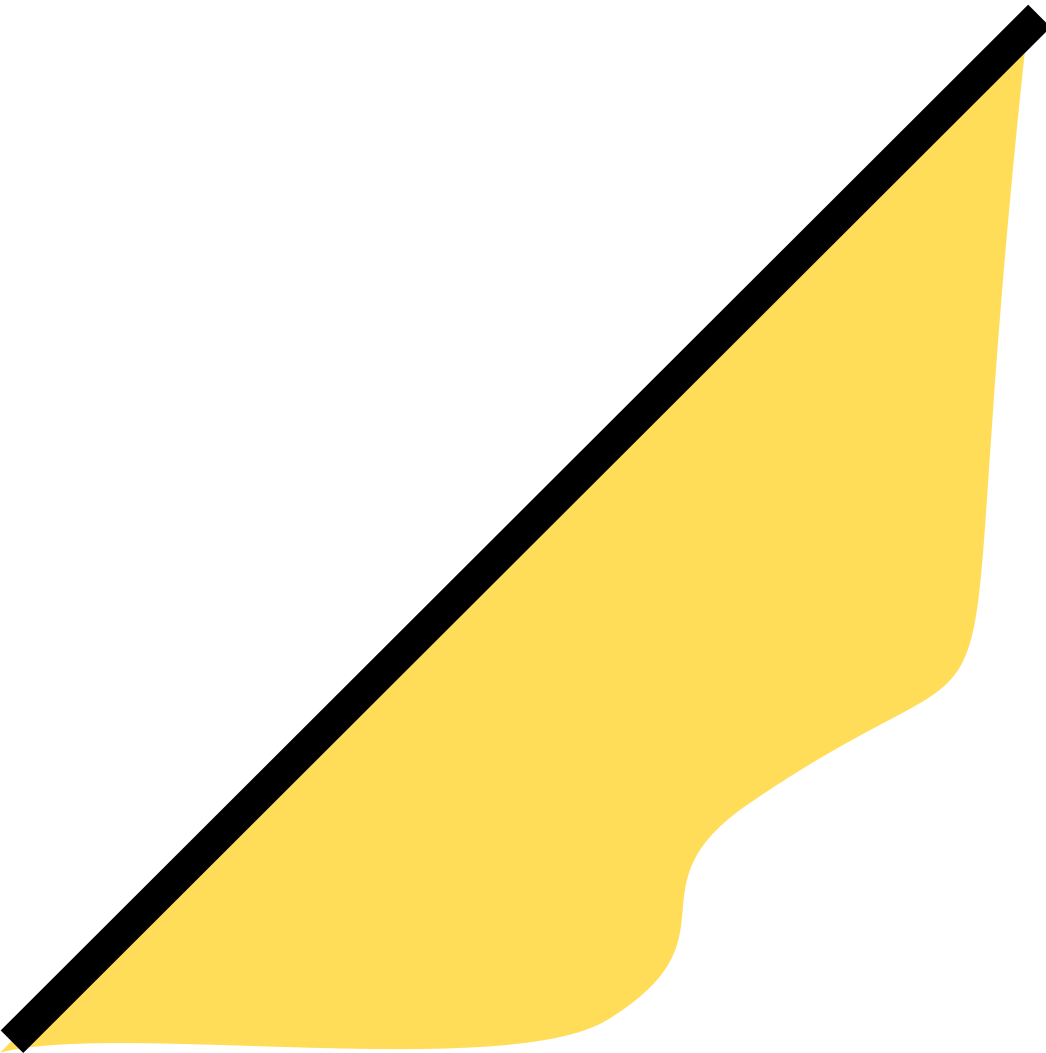}}\right)=-\text{sgn} \left( \raisebox{-5pt}{\includegraphics[height=0.5cm]{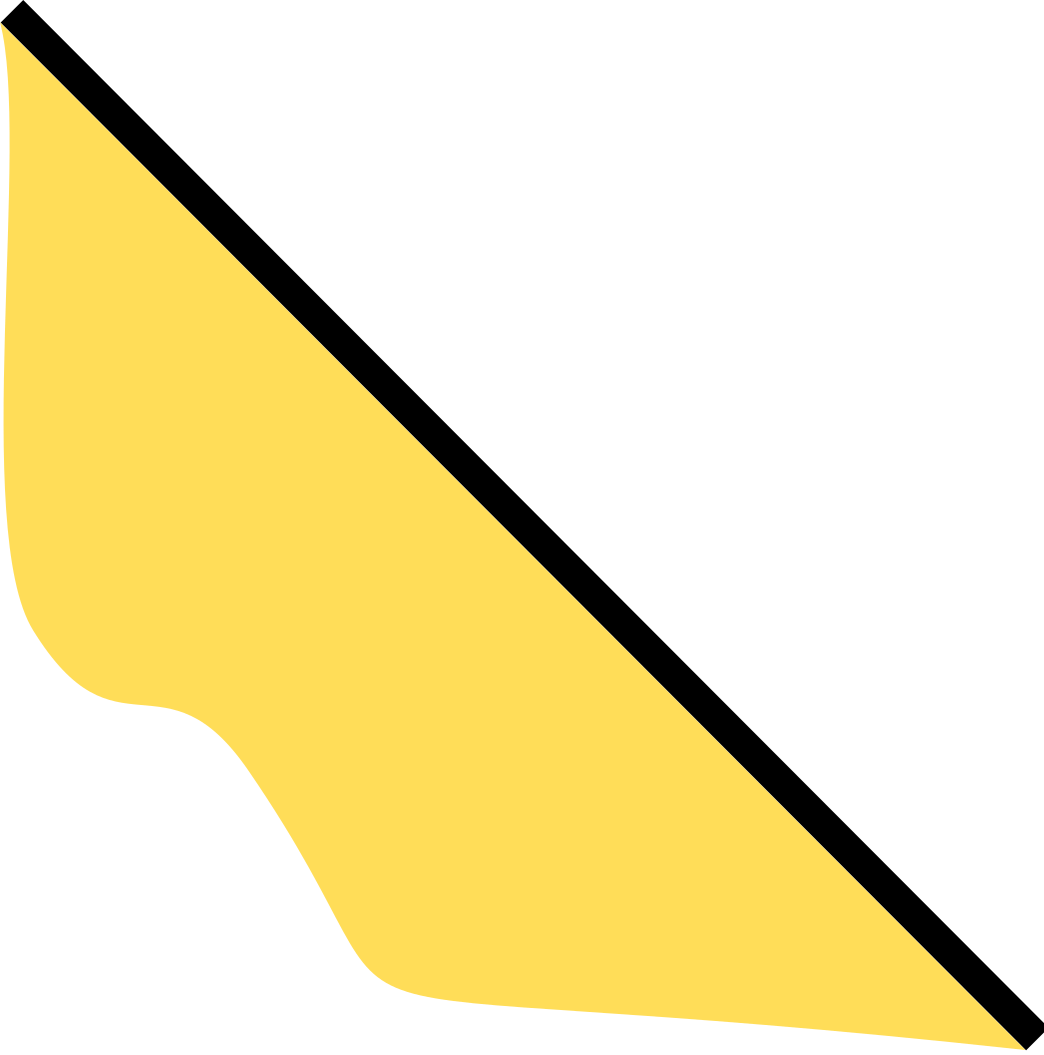}} \right)   =1.$$

\item $j_i$ are the codimension-two junctions between boundary components. For those joints that are formed by at least one null boundary, the form of the integrand in \eqref{masterformula} is given by

\begin{equation}
\alpha_{j_i} = 
\begin{cases}
\log |k_\mu n^\mu| \\
\log |k_\mu s^\mu|\\
\log |\frac{1}{2}k_\mu \bar{k}^\mu|,
\end{cases}
\end{equation}

where $k^\mu$ and $\bar{k}^\mu$ are taken to be the future directed vectors tangent to the null surfaces and $n^\mu$ ($s^\mu$) is the future-directed (outward-directed) unit normal to the spacelike (timelike) surface. For such terms the signs $\sgn(j_i)$ are simply given by the product of the surfaces signs. Joints that are formed only by spacelike and timelike boundaries have a different set of rules that we will not cover here as they do not appear on WdW patches (see \cite{Poisson} for a full discussion of all possible joint actions).
\end{itemize}

Although the terms described so far are `canonical', in the sense that they lead to a well-defined variational principle, it remains a  notorious dependence on the parametrization of null boundaries. In order to cancel this parametrization dependence, it is possible to add an extra countertem depending on the expansion of codimension-two sections along the null boundaries
\begin{equation}
\label{expansion}
\Theta = \partial_\lambda \,\log\,\sqrt{\gamma}
\;. 
\end{equation}
 We shall refer to this addition as the  expansion counterterm:
\begin{equation}
\label{Ict}
I_{\Theta} = \sum\limits_i\dfrac{\text{sgn}(N_i)}{8\pi G} \int\limits_{N_i}  \text{d}\lambda\, \text{d}^{d}x \;\sqrt{\gamma}\; \Theta \;\log(\ell_\Theta\, \cdot |\Theta|).
\end{equation}

The appearance of the new length scale $\ell_\Theta$ is interesting. It represents a qualitatively new feature of AC complexity which activates itself  precisely in cases where the entropy has a dynamical behavior and the effective Hilbert space supporting the singularity  changes its dimension. 
The presence of $I_\Theta$ has been regarded as necessary to  guarantee the positivity of complexity \cite{Ross} as well as the correct black hole complexity dynamics from collapsing geometries and the verification of the switchback effect \cite{VaidyaI, VaidyaII, WhichAction}. Its precise meaning in microscopic treatments inspired by the notions of circuit complexity remains quite mysterious (cf. \cite{JeffersonMyers, Myersmixed, Myerscoherent, MyersTFD }).

\section{Expanding entropy dominance of AC}
\label{sec:terminals}

As emphasized in the introduction, the  asymptotic behavior of the AC complexity for an eternal black hole is completely decoupled  from that of the entropy. In this section we show that this decoupling does not hold when the entropy has a strong dynamical component.   The primary example is that of a singularity inside an expanding bubble embedded in an ambient AdS spacetime. The singularity eventually crunches the whole AdS spacetime in a finite time, as measured by the asymptotic global time. The boundary of the bubble has an acceleration horizon which serves as the boundary of the past causal domain $D^- ({\cal S}^*)$. As a result, the entropy of this crunch singularity is infinite. 

\begin{figure}[H]
$$\includegraphics[width=8cm]{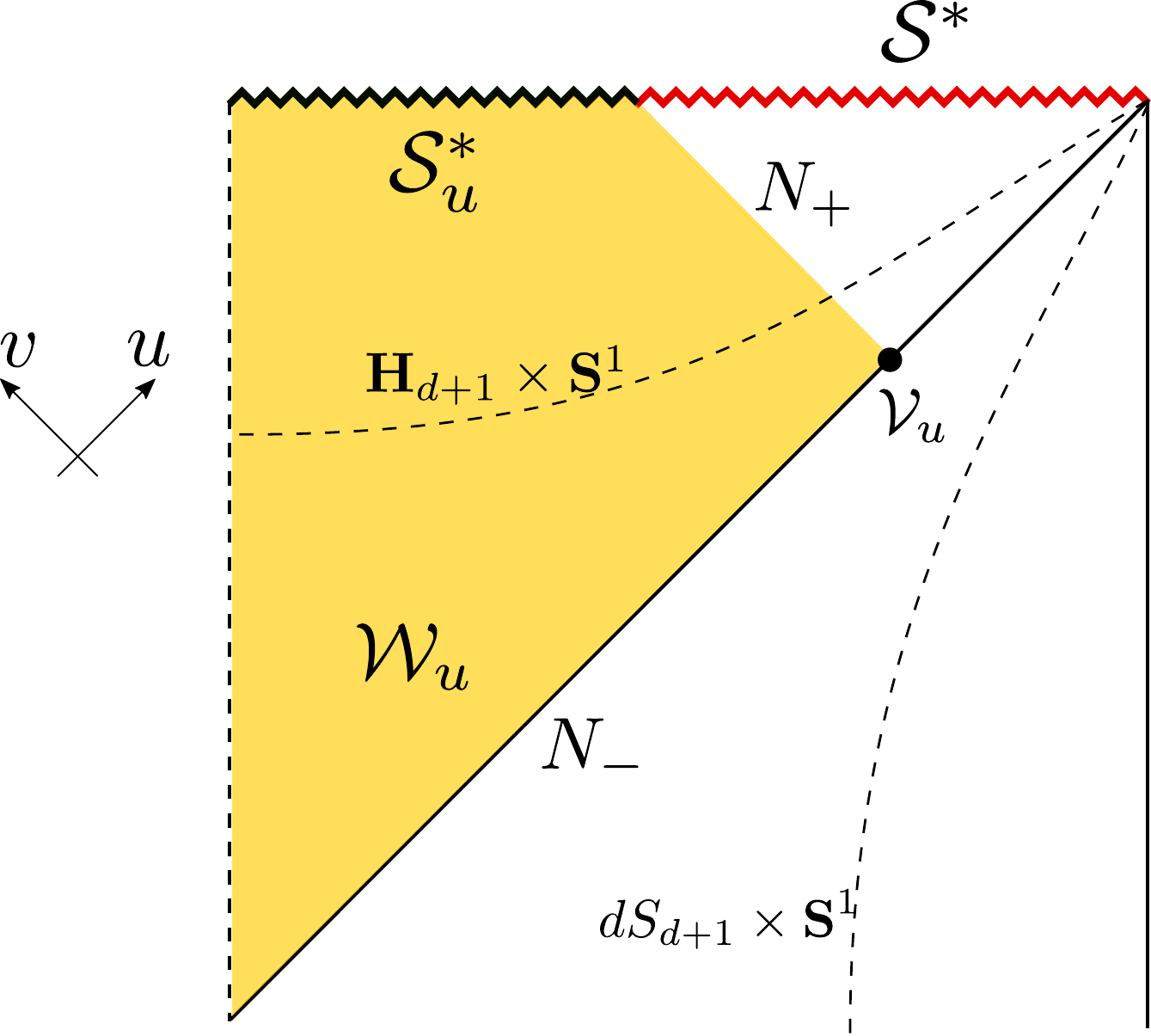} $$   
\begin{center}
\caption{\emph{ The causal structure of the topological crunch singularity and its associated WdW patch. }
\label{fig:tc}}
 \end{center}
\end{figure}

In order to rely on analytic methods, we first consider the `topological cruch' model (cf. \cite{Banados, maldapim, BarbonRabinoCrunches}),   which describes a time-dependent compactification of pure AdS$_{d+3}$ with topology AdS$_{d+2} \times {\bf S}^1$, where the ${\bf S}^1$ shrinks to zero size in finite boundary time, producing a spacelike singularity in the interior as shown in Figure \ref{fig:tc}. A holographic interpretation of this model uses a CFT$_{d+2}$ on a spatial manifold with topology $ {\bf S}^{d} \times {\bf S}^1$, where  the sphere is static and the circle shrinks to zero size in finite time. A conformally related description is that of the same CFT on a fixed-size circle, times a de Sitter spacetime.

In the FLRW patch of the $AdS_{d+2}$ the metric for this model is given by
\begin{equation}
\text{d}s^2 = -\text{d}\tilde{t}^{\,2} +  \sin^2 (\tilde{t}\,) \,\text{d}{\bf H}_{d+1}^2 + \cos^2 (\tilde{t}\,) \,\text{d}\phi^2 \, ,
\end{equation}
where we have set the AdS scale to unity and also consider that to be the curvature scale of the boundary metric. The element $d{\bf H}_{d+1}^2$ stands for the unit metric on the $(d+1)$-dimensional Euclidean hyperboloid ,whereas the coordinate $\phi$ is the angle that parametrizes the compact circle ${\bf S}^1$. The time $\tilde{t}$ foliates the cosmology with sections of topology ${\bf H}_{d+1} \times {\bf S}^1$, producing a singularity $\mathcal{S}^*$ when the circle shrinks to zero size at time $\tilde{t}_*=\pi/2$.

In order to perform the computation of the action, we will work in null Kruskal compact coordinates given by

\begin{eqnarray}
\tan u=  e^{\eta + \chi} \;, \qquad \tan v = e^{\eta - \chi} \, ,
\end{eqnarray}
where $\chi$ is the usual radial coordinate in ${\bf H}_{d+1}$ and we have defined the conformal time over the AdS$_{d+2}$ as  $\eta =2\tan^{-1} \left( \exp(\tilde{t}\,)\right)$. Performing the coordinate changes, we get the metric

\begin{equation}
\label{topmetric}
\text{d}s^2 =  \sec ^2(u-v) \left[ -4 \,\text{d}u \text{d}v + \sin ^2(u-v) \,\text{d}\Omega^2_{d} + \cos^2 (u+v)\, \text{d} \phi^2 \right].
\end{equation}

Considering now the set of nested WdW patches $\CW_u$ labeled by the null coordinate $u$, we are ready to calculate the different contributions to the on shell action from the prescription \eqref{masterformula}. As we will only care about asymptotic behaviors and not the exact full action, it will suffice to consider stripes $\delta \CW_u$ of thickness $\delta u$ for which computations render sensibly simpler results. The bulk action piece of such stripe is given by

\begin{eqnarray}
I_{\text{bulk}}[\delta \CW_u] &=& -\dfrac{(d+2)}{8 \pi G} \int\limits_{\delta \CW_u} \text{d}^{d+3}x \, \sqrt{-g} \\
&=& -\dfrac{2(d+2)V_\Sigma}{8 \pi G} \int\limits^{u+\delta u}_{u} \text{d}u  \int\limits_0^{\pi/2-u} \text{d}v \; \dfrac{\tan^d(u-v)}{\cos^3(u-v)}\cos(u+v)\,.
\end{eqnarray}
where $V_\Sigma$ stands for the area of the ${\bf S}^{d}\times {\bf S }^1$ manifold. Performing this integral and expanding around $u\sim u_*=\pi/2$ we can obtain the asymptotic limit for the bulk action growth

\begin{equation}
\dfrac{\text{d} }{\text{d}u} I_{\text{bulk}}[ \CW_u] \approx -\dfrac{V_\Sigma}{8 \pi G} \,\dfrac{2(d+2^{-d-1})}{d+1}\left( \dfrac{1}{u_*-u} \right)^{d+1}.
\end{equation}

In order to calculate the codimension-one boundary terms, we will choose to parametrize all null boundaries affinely, so that the extrinsic curvature vanishes and such contributions are identically zero. The only non-trivial YGH contribution will be that of the spacelike boundary at the singularity, which is located at $\tau \equiv v+u =\pi /2$. This term is given by
\begin{equation}
I_{\text{YGH}}[ \CW_u] = -\dfrac{1}{8 \pi G} \int\limits_{\CS^*_u} \text{d}^{d+2}x \,\sqrt{\gamma}\,K \, ,
\end{equation} 
where $\gamma$ is the induced metric on $\CS^*_u$ and the integrand is calculated using $\sqrt{\gamma} \, K = g_{\tau \tau}^{-1/2} \, \partial_\tau \sqrt{\gamma}$. Performing the integral again for the slab of thickness $\delta u$ we get the growth rate and its late time limit
\begin{equation}
\dfrac{\text{d} }{\text{d}u}I_{\text{YGH}}[ \CW_u]  = \dfrac{V_\Sigma}{8 \pi G} \dfrac{\tan^{d+1}(2u)}{\sin^{}(2u)}\approx \dfrac{V_\Sigma}{8 \pi G} 2^{-d-1}\left( \dfrac{1}{u_*-u} \right)^{d+1}.
\end{equation}
The only codimension-one contribution that is now left is that of the expansion counterterms \eqref{Ict}. Let us consider thus the WdW null boundaries, which will be given by constant $u,v$ hypersurfaces. For instance, we may start with the past boundary, given by the surface $N_- \equiv (u,0,\Omega_0, \phi_0)$, where the coordinate $u$ here will parametrize the geodesic. Introducing this curve into the geodesic equation, however, we can see that such parametrization is not affine, but rather has the following surface gravity
\begin{equation}
\kappa(u) = 2 \tan u \,.
\end{equation}
Following the standard procedure, we can find an affine parameter $\lambda_-$ from the relation
\begin{equation}
\dfrac{\text{d} \lambda_-}{\text{d}u} = \exp{\int\limits_0^u\kappa(\sigma) \, \text{d} \sigma}\, ,
\end{equation}
which for our case yields
\begin{equation}
\lambda_- = \dfrac{1}{\alpha_-} \tan u\,,
\end{equation}
and we have introduced a constant $\alpha_-$ that parametrizes the freedom to shift the affine parameter. From \eqref{topmetric} we can extract the determinant of the induced transverse metric as well as its expansion and express them in terms of $\lambda_-$
\begin{eqnarray}
\sqrt{\gamma} &=& (\alpha_-\lambda_-)^{d}\, ,\\
\Theta &=&  \dfrac{d}{\lambda_-}\, .
\end{eqnarray}
Feeding this into the counterterm definition \eqref{Ict} we get the contribution to the action
\begin{eqnarray}
\label{counterU}
I^-_{\Theta} &=& \dfrac{V_\Sigma}{8 \pi G} \int\limits_0^{\frac{1}{\alpha_-}\tan u}  \text{d}\lambda_- \, \alpha_- \,(\alpha_-\lambda_-)^{d-1}d \, \log \left(\LL \dfrac{d}{\lambda_-} \right) \\  
&=&\dfrac{V_\Sigma}{8 \pi d G} \tan ^{d}u \, (d \, \log (d \, \LL \, \alpha_- \cot u)+1)\,.
\end{eqnarray}
A similar procedure gives us the affine parameter for the future null boundary $N_+$
\begin{equation}
\lambda_+ = \dfrac{1}{\alpha_+} \tan (u-v)\, ,
\end{equation}
which yields
\begin{equation}
\sqrt{\gamma} = (\alpha_+   \lambda_+ )^{d} (\alpha_+  \lambda_+  \sin (2 u)+\cos (2 u)),
\end{equation}
\begin{equation}
\Theta = \frac{1}{\lambda_+ }\left(d+1-\frac{1}{\alpha_+  \lambda_+  \tan (2 u)+1} \right),
\end{equation}
and we can calculate the corresponding counterterm \footnote{Although this integral is analytically solvable, the result is rather cumbersome and not particularly illuminating. We omit therefore such explicit expression since we will only care about its late time expansion.}

\begin{eqnarray}
\label{counterV}
\scriptsize
I^+_{\Theta} &=&  -\dfrac{V_\Sigma}{8 \pi G} \int\limits^{\frac{1}{\alpha_+} \tan(2u-\frac{\pi}{2})}_{\frac{1}{\alpha_+} \tan u} \text{d}\lambda_+ \, \text{d}^{d+1}x \, \sqrt{\gamma} \,\Theta \,\log(\LL \,|\Theta|).
\end{eqnarray}
Finally, we must also calculate the contribution from the codimension-two joints of $\CW_u$. As the joint $N_+ \cap \CS^*_u $ has vanishing volume, the only one that will produce a non-trivial contribution is $\CV_u = N_- \cap N_+$. Following the rules on section \ref{sec:action}, the contribution from this joint is given by  
\begin{equation}
\label{Sjoints}
I_{\CV_u} =  -\dfrac{1}{8 \pi G} \oint\limits_{\CV_u} \text{d}^{d+1}x \, \sqrt{\sigma} \, \log | \tfrac{1}{2}k_+ \cdot k_- |\,  ,
\end{equation}
where $k_+$ and $k_-$ are respectively the null tangent vectors of the future and past boundaries. We choose these vectors to be
\begin{eqnarray}
k_+ &=& \alpha_+ (dT+dX)\,,\\
k_- &=& \alpha_- (dT-dX)\,,
\end{eqnarray}
so that they satisfy $k_{\pm} \cdot \partial_T=\alpha_{\pm}$ and $\alpha_{\pm}$ are normalization constants. Substituting these values into \eqref{Sjoints} we get the contribution of the joint
\begin{eqnarray}
\label{jointint}
I_{\CV_u} &=& -\dfrac{1}{8 \pi G} \log \left(\alpha_+ \, \alpha_-   \dfrac{\cos^2u}{2}\right) \oint \text{d}^{d+1}x \, \sqrt{\sigma}  \\ &=& \nonumber -\dfrac{V_\Sigma }{8 \pi G}\log \left(\alpha_+ \, \alpha_-   \dfrac{\cos^2u}{2}\right)  \tan^{d} u\, .
\end{eqnarray}

We observe that the dependence on $\alpha_\pm$ cancels out when \eqref{counterU}, \eqref{counterV} and \eqref{jointint} are added up, and accordingly we can set them to 1 in these expressions. Collecting all results for a late time expansion $u \sim u_*$, we get the following  behavior for the contributions
\begin{eqnarray}
I_{\text{bulk}} &\approx & -\dfrac{S}{\pi d} \left( \dfrac{d+2^{-d-1}}{d+1}\right)\, , \\ 
I_{\text{YGH}} &\approx & \dfrac{S}{\pi d} \times 2^{-d-1}\, ,\\
I^-_{\Theta}  &\approx & \dfrac{S}{2\pi d}\left[ 1+ \log\left(\dfrac{V_\Sigma}{4G} \right)+d\log\left(d\, \LL\right)-\log S\right], \\
I^+_{\Theta}  &\approx & \dfrac{S}{2\pi d} \left[ f(d) + \log\left(\dfrac{V_\Sigma}{4G} \right)+d\log\left(d\,\LL\right)-\log S \right], \\
I_{\CV_u}  &\approx & \dfrac{S}{2 \pi d} \left[d\log(2)- 2 \log\left(\dfrac{V_\Sigma}{4G} \right)  +2 \log S \right], 
\end{eqnarray}
where we are dropping terms in fractional powers of the entropy $S$ along the horizon, defined as  
\begin{eqnarray}
\label{entropytop}
S &=& \dfrac{1}{4 G} \oint  \text{d}^{d+1}x \, \sqrt{\sigma}  \\ &=& \nonumber \dfrac{V_\Sigma }{4 G} \tan^{d} u \approx \dfrac{V_\Sigma}{4 G}\left( \dfrac{1}{u_*-u} \right)^{d},
\end{eqnarray}
and keeping only the leading and next-to-leading contributions to the action. The coefficient $f(d)$ is an $\CO(1)$ positive constant given by
$$\scriptstyle  (d+1)f(d)=  \left(d^2+d-1\right) \, _2F_1\left(1,d;d+1;2+\frac{2}{d}\right)-d^2+2 (d+1) d \left(\coth ^{-1}(d+1)-\, _2F_1\left(1,d+1;d+2;2+\frac{2}{d}\right)\right)+2^{-d} \, _2F_1\left(1,d;d+1;1+\frac{1}{d}\right).$$

After adding up all contributions, we can see that the $S\log S$ divergence cancels out, yielding a total late-time complexity dynamics linear in the entropy

\begin{equation}
 \CC^* \approx   a\, S
\end{equation}
with
\begin{equation}
a= \dfrac{1}{\pi d} \left[\dfrac{-d-2^{-d-1}}{d+1} +  2^{-d-1} + \dfrac{1}{2} \left(1+f(d)+d\log (2) \right) +d\log \left( d\, \LL\right) \right].
\end{equation}

As we see, the complexity growth is fully controlled by that of the entropy, diverging as measured by any null coordinate along the horizon. The sign of such growth however will depend on the coefficient, which is essentially controlled by the size of $\LL$,  yielding a positive rate when $\LL \gtrsim 1$ for any dimension.\\

\begin{figure}[h]
$$\includegraphics[width=5cm]{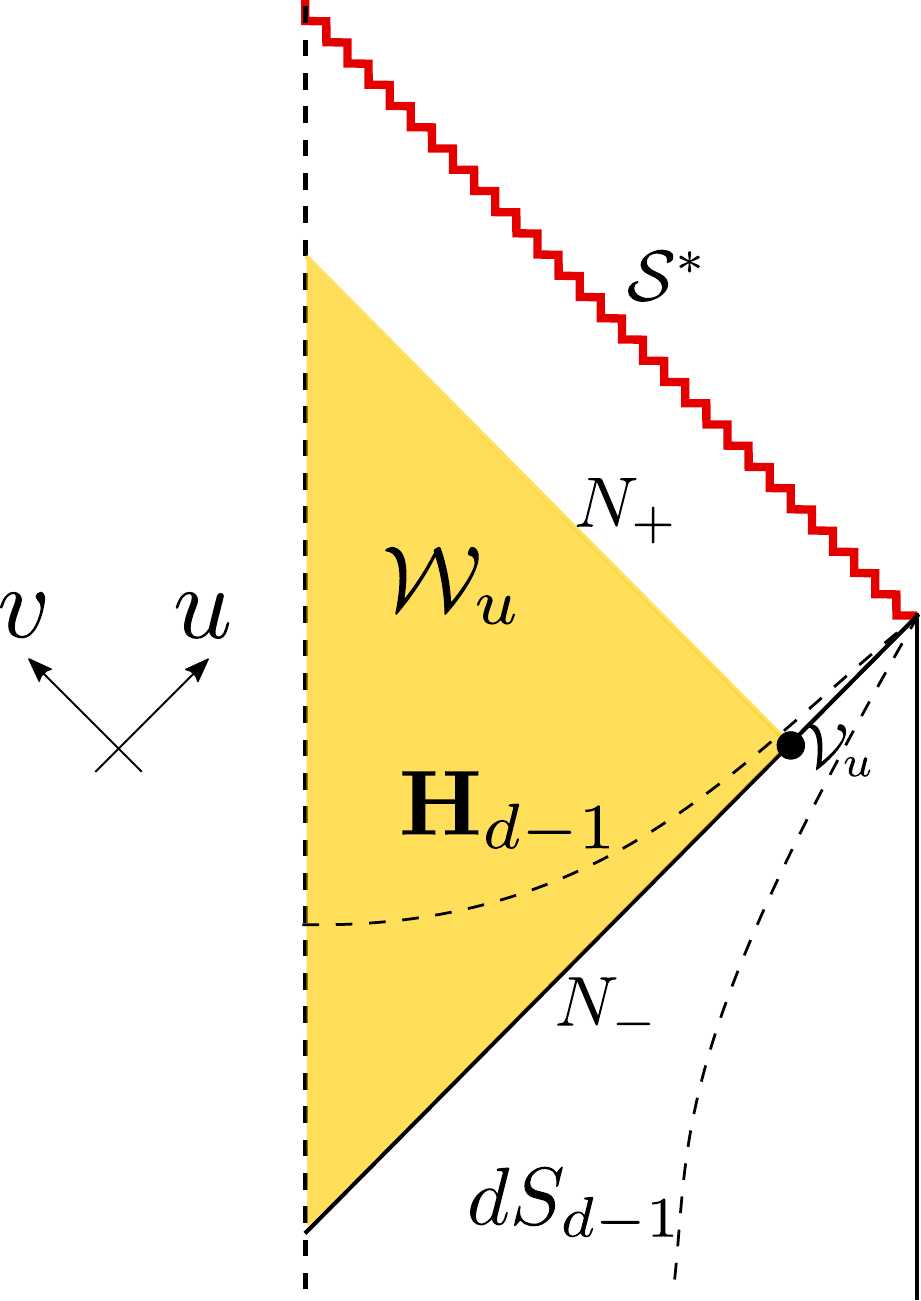} $$   
\begin{center}
\caption{\emph{ The idealized  state generating a null singularity by the collision of a thin-walled bubble with dS$_{d+1}$ worldvolume. Notice that WdW patches anchored at the horizon do not touch the singularity. }\label{fig:tw}}
\label{fig:thinwall}
\end{center}
\end{figure}

It would be interesting to generalize this result to more general solutions with expanding horizons. The need to consider scalar fields with non-trivial potentials generally prevents us from a completely analytic treatment. However, we can offer evidence that the result found is quite robust by examining a similar situation in the so-called thin-wall approximation. Suppose that the bubble has a very narrow outer shell, so that we can describe it as a thin wall expanding into AdS$_{d+2}$, with a de Sitter induced metric. In this case the singularity can be regarded as a null future-directed surface emerging from the boundary  impact time at $t=t_*=\pi $  (see Figure \ref{fig:thinwall}). 

The formal analysis is very similar to that of the topological crunch with the difference that the shrinking ${\bf S}^1$ is not present now. The bulk metric can be obtained therefore simply removing the $\text{d} \phi^2$ factor in \eqref{topmetric} 
\begin{equation}
\label{thinmetric}
\text{d}s^2 =  \sec ^2(u-v) \left[ -4 \, \text{d}u \text{d}v + \sin ^2(u-v) \, \text{d}\Omega^2_{d} \right],
\end{equation}
and the calculations follow very easily from the previous ones. In effect, the bulk contribution is obtained as

\begin{eqnarray}
I_{\text{bulk}}[\CW_u] &=&  -\dfrac{(d+1)V_\Omega}{8 \pi G} \int\limits^{u}_{0} \text{d}u' \int\limits^{u'}_{0} \text{d}v \; \dfrac{\tan^d(u'-v)}{\cos^3(u'-v)} \approx \dfrac{V_\Omega}{8 \pi G d} \left( \dfrac{1}{u_*-u} \right)^{d}.
\end{eqnarray}
where as usual we have performed an expansion around $u \sim u_*$ in the last equality. As the timelike surface at $u=v$ has vanishing induced metric, all codimension-one boundaries of this WdW patch vanish when the parametrization is taken to be affine along the null ones $N_{\pm}$. We need however to compute the countertems for the later. For both boundaries we get the quantitites \footnote{We omit here the normalization constants $\alpha_\pm$ as its cancellation is analogous to the previous case}
\begin{eqnarray}
\lambda_\pm &=&  \tan (u-v)\,, \\
\sqrt{\gamma}_{\pm} &=& (\lambda_\pm)^{d}\,, \\
\Theta_\pm &=& \dfrac{d-1}{\lambda_\pm}\,.
\end{eqnarray}
And the counterterms are very similar to \eqref{counterU}
\begin{eqnarray}
\label{TWct}
I^-_{\Theta} = I^+_{\Theta} &=& \dfrac{V_\Omega}{8 \pi G} \int\limits_0^{\tan u}  \text{d}\lambda \,  \lambda^{d-1} d \, \log \left(\LL\, \dfrac{d}{\lambda} \right) \\  
&=&\dfrac{ V_\Omega}{8 \pi d G} \tan^{d}u \, (d \, \log (d \, \LL \cot u)+1), \\
 &\approx& \dfrac{S}{2\pi d} \left(1+\log \left( \dfrac{V_\Omega}{4G}\right)+ d\log \left( d \,\LL\right) -\log S \right) ,
\end{eqnarray}
where we substituted again the entropy along the horizon, which remains identical as in the previous example \eqref{entropytop}. The joint contribution will also have the same form as in the previous example, therefore cancelling again the $S \log S$ leading divergence in $I_{\Theta}^{\pm}$ for the asymptotic limit $u\sim u_*$. Adding up all pieces of the action we may obtain the total complexity
\begin{eqnarray}
\CC^*\approx  a \,S \,,
\end{eqnarray}
with
\begin{eqnarray}
a= \dfrac{1}{\pi d} \left[-1+\dfrac{1}{2}d\log(2)  + d\log \left( d \,\LL \right)\right],
\end{eqnarray}
an expression that tells us again that the leading divergence is guaranteed to be positive as long as the counterterm scale satisfies $\LL \gtrsim 1$.

\begin{figure}[h]
$$\includegraphics[width=4cm]{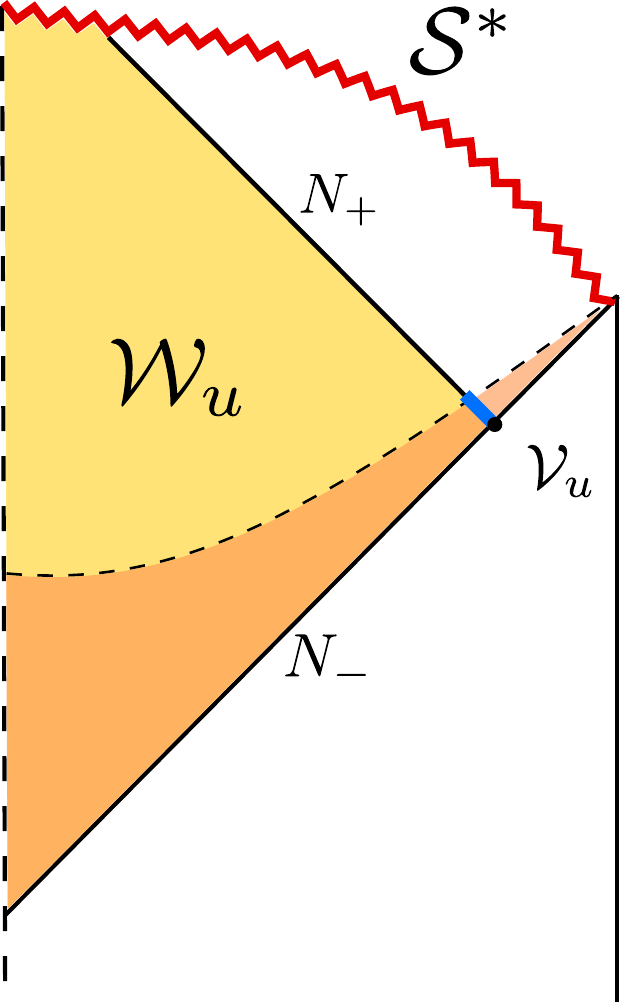} $$   
\begin{center}
\caption{\emph{ Maximally extended FLRW solution embedded in AdS space. The early-times approximation $a(\tilde{t}\,)\approx \tilde{t}$ guarantees that the geometry in the vecinity of $N_-$ (shaded region) is the same as in the pure AdS solutions, implying that $I^-_\Theta$ and $I_{\mathcal{V}_u}$ are identical to those of the vacuum cases. As the same is true for the near-horizon piece of $I^+_\Theta$ (in blue), the logarithmic divergence is generically cancelled provided no new dominating contributions arise from the deep interior piece of $I^+_\Theta$.}}
\label{fig:thickwall}
\end{center}
\end{figure}

As both the latter examples come from spacetimes that are locally AdS, one might be suspicious about the seemingly miraculous cancellation of the logarithmic divergence. Nonetheless, there are reasons to think that such cancellation is generic for more realistic solutions involving a backreacting scalar field that drives a homogeneous cosmology behind the horizon. Indeed, for any negatively curved FLRW solution to enjoy a smooth horizon, the behavior of the scale factor around the time origin must be of the form $a(\tilde{t}\,) \approx \tilde{t}$. This scaling coincides with the structure present in the near-horizon region from the previous examples and therefore will yield the very same contributions for the $I_\Theta^-$ and $I_{\CV_u}$ terms. The future null counterterm $I^+_\Theta$ on the other hand, will pick up information from the full cosmological solution which in turn will depend on the details of the particular model. In the two analytic examples above, however, it is easy to see that the integrals \eqref{counterV} and \eqref{TWct} are strongly dominated by the near-horizon region, the one responsible of the logarithmic divergence. The far interior, on the other hand, contributes at most linearly to the complexity. As the cosmological solution must be that of the thin wall approximation for a neighborhood around the horizon, the logarithmic contribution from $I^+_\Theta$ will remain identical and accordingly will produce the generic cancellation of the leading divergence (cf. Figure \ref{fig:thickwall}). This of course does not prevent a possible restoration of higher than linear divergencies in both the bulk and  $I^+_\Theta$ contributions when the full cosmological dynamics is taken into account. The estimation of such effects is beyond the scope of this paper.

\section{The local case}

\noindent

\begin{figure}[h]
$$\includegraphics[width=9cm]{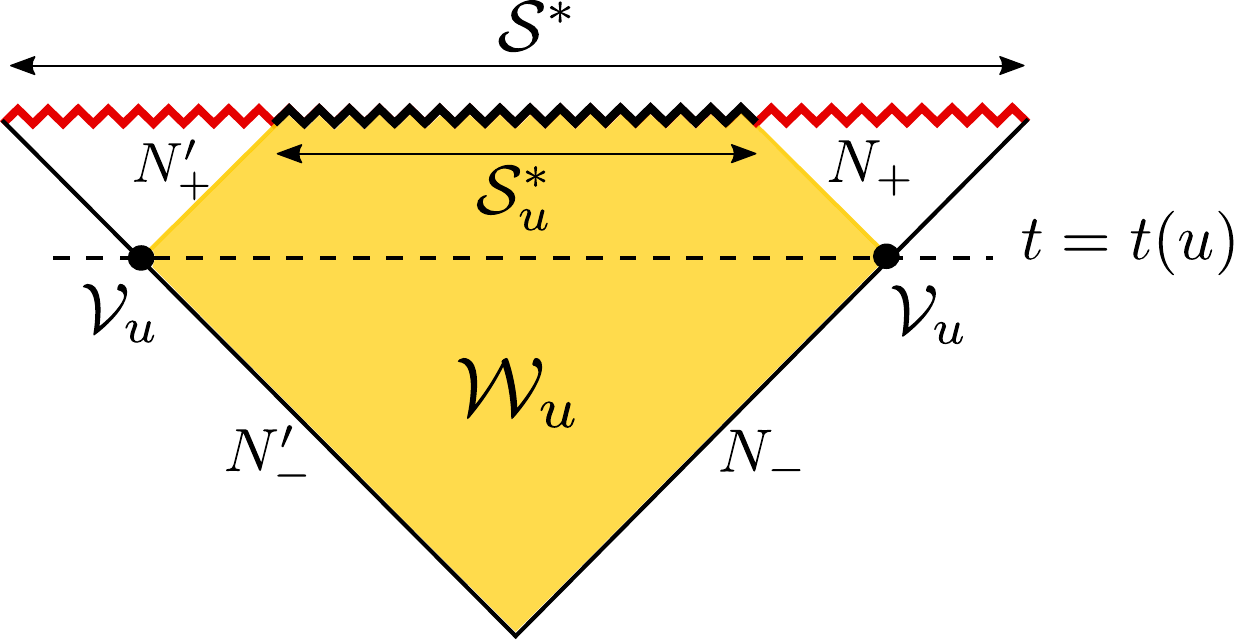} $$   
\begin{center}
\caption{\emph{ WdW patch for a Kasner slab ${\cal S}^*_u = \left[-{u \over 2} , {u\over 2} \right] \times {\bf R}^{d-1}$. }}
\label{fig:kasner}
\end{center}
\end{figure}

The definition of terminal complexity is quasilocal in the sense that it makes no explicit reference to timelike boundaries which may support standard CFT duals. In particular, we can define it for any subset of a spacelike singularity. In the limit of a small patch, it is known (cf. \cite{misner, BKL,BKL2,BKL3}) that the metric near a GR singularity is well approximated by the Kasner form
\begin{equation}
\label{kas}
ds^2 = -dt^2 + \sum_{j=1}^{d+1} (H\,t)^{2p_j} dx_j^2\;,
\end{equation}
with parameters $p_j$ satisfying $\sum_j p_j = \sum_j p_j^2 =1$ and $x_j$ restricted to a small domain $H\Delta x_j \ll 1$. In order to adapt the discussion to the symmetries of the Kasner metric, we  pick one spatial coordinate, say $x_1$, and define the singular set to be a `slab' ${\cal S}^* = \left[-{u_* \over 2} , {u_*\over 2} \right] \times {\bf R}^{d-1}$ for some finite $u_*$ measuring the $x_1$ coordinate length of the finite interval. The set of WdW patches indicated in Figure \ref{fig:kasner} intersect ${\cal S}^*$ along the nested  `slabs' ${\cal S}^*_u =   \left[-{u \over 2} , {u\over 2} \right] \times {\bf R}^{d-1}$. 

In this construction, we regard the holographic data as specified on ${\cal V}_u = \{-{u\over 2}\} \times {\bf R}^{d-1} \cup  \{{u\over 2}\} \times {\bf R}^{d-1}$, and the $x_1$ coordinate takes the role of `holographic' emergent direction. The fact that the holographic data lies on disconnected spaces, interpolated by the `bulk' WdW patches, makes this construction similar to the standard eternal AdS black hole and its dual product CFTs \cite{Maldacenaeternal}, with the crucial difference that here the entropy density 
\begin{equation}
{\tilde S} = {1\over V_{{\bf R}^{d-1}}} {{\rm Vol} ({\cal V}_u )\over 4G} \, ,
\end{equation}
vanishes as  the singularity is approached in the limit $u\rightarrow u_*$. Properly speaking, we will consider the geometric description given by the Kasner metric to be appropriate only up to a Planckian cutoff away from the singularity. Only in this setup the semiclassical entropy defined here is reliable and $S \gg 1$ is guaranteed. Our formal notion of `vanishing' entropy is to be understood therefore in this approximate sense, as the strict $u\rightarrow u_*$ limit is out of our effective theory.

Let us pick spacetime units so that the `Hubble rate' $H=1$, and pass to conformal coordinates 
\begin{equation}
\label{kasnermetric}
\text{d}s^2 = \left((1-p_1) \tau \right)^{\frac{2p_1}{1-p_1}} \left(-\text{d}\tau^2 + \text{d}x_1^2 +  \sum\limits_{i=2}^{d+1}\left((1-p_1)\tau\right)^{\frac{2(p_i-p_1)}{1-p_1}} \text{d}x_i^2 \right),
\end{equation} 
where $\tau$ is the conformal time in the $(t, x_1)$ plane. As in this case we are dealing with a Ricci flat solution, the bulk contribution of $\CW_u$ will be trivially zero. It will suffice thus to compute the expansion counterterms, joints and the YGH contribution from the singular locus. For the latter, it is easy to see from \eqref{kasnermetric} that $\sqrt{\gamma}\,K=1$, implying the very simple contribution
\begin{equation}
I_{\text{YGH}}= \dfrac{V_{{\bf R}^{d-1}}}{4\pi G}\, u \,,
\end{equation}
where $V_{{\bf R}^{d-1}}$ is the volume of the non-compact directions, appearing here in the sense that we may define a finite complexity density $\widetilde{\CC}=\CC /V_{{\bf R}^{d-1}}$. In order to compute the counterterms, we may define as usual the null coordinates by $\tau, x_1 = \frac{1}{2} (u\pm v)$. Starting with the past boundary $N_- = (u,v_0,\vec x_{i,0}),$ we find that its surface gravity is given by

\begin{equation}
\kappa = -\frac{2 p_1}{(p_1-1) (u+v_0)}\, ,
\end{equation}
and following the standard procedure we get the correct affine parameter
\begin{equation}
\lambda_-(u) =  \dfrac{1}{\alpha_-}(u+v_0)^{1/\delta}\, ,
\end{equation}
where we defined for simplicity $\delta= \frac{1-p_1}{1+p_1}$. We can calculate now the induced volume element and the expansion
\begin{eqnarray}
\sqrt{\gamma} &=& -\dfrac{\delta}{1+\delta}(\alpha_- \lambda_-)^\delta\,, \\
\Theta &=& \dfrac{\delta}{\lambda_-}\,.
\end{eqnarray}
Feeding it into the couterterm action we get
\begin{eqnarray}
I_{\Theta}^- &=& -\dfrac{ V_{{\bf R}^{d-1}}}{8 \pi G} \int\limits_{\lambda_-(u_0)}^{\lambda_-(u)} \text{d} \lambda_- \, \dfrac{\delta^{2}}{\delta+1} \,\alpha_-^\delta \, \lambda^{\delta-1} \, \log\left(\dfrac{\LL \,\delta}{\lambda_-} \right) \\  
&=& -\dfrac{ V_{{\bf R}^{d-1}}}{8 \pi G} \dfrac{1}{\delta+1}(u+v_0)\left[   1+ \delta \log \left(\delta \, \LL \alpha_- \right) -\log \left(u+v_0 \right)  \right]  - (u \leftrightarrow u_0) .\nonumber
\end{eqnarray}

Repeating the procedure above for the future null boundary $N_+$ we can see that the counterterm contribution is given by an almost identical expression where we must exchange the roles $u, u_0 \leftrightarrow u$ and $v_{0} \leftrightarrow v_{0}, v_*$ with $v_*=-v_0=-u_0=u_*$ the value at the singularity.
\begin{eqnarray}
I_{\Theta}^+ &=& \dfrac{ V_{{\bf R}^{d-1}}}{8 \pi G} \int\limits^{\lambda_+(v_*)}_{\lambda_+(v_0)} \text{d} \lambda_+ \, \dfrac{\delta^{}}{\delta+1} \,\alpha_+^\delta \, \lambda_+^{\delta-1} \, \log\left(\dfrac{\LL \,\delta}{\lambda_+} \right) \\  
&=& -\dfrac{ V_{{\bf R}^{d-1}}}{8 \pi G} \dfrac{1}{\delta+1}(u+v_0) \left[  1+ \delta\,\log \left( \delta \LL \alpha_+  \right) -\log \left(u+v_0 \right) \right] .\nonumber
\end{eqnarray}
Now, we must calculate the contribution from the joints of these surfaces. Choosing the normal vectors
\begin{eqnarray}
k_+ &=& \alpha_+ (d\tau+dx_1)\,,\\
k_- &=& \alpha_- (d\tau-dx_1)\,,
\end{eqnarray}
we get the contribution
\begin{equation}
I_{\CV_u}= \dfrac{ V_{{\bf R}^{d-1}}}{4 \pi G} \dfrac{\delta}{\delta+1}\left(u+v_0\right) \log \left( \alpha_+ \, \alpha_- \,\left((u+v_0)\dfrac{\delta}{\delta+1} \right)^{\frac{\delta-1}{\delta}} \right)\,,
\end{equation}
and similarly for the south tip
\begin{equation}
I_{\southcorner}= -\dfrac{ V_{{\bf R}^{d-1}}}{8 \pi G} \dfrac{\delta}{\delta+1}\left(u_0+v_0\right) \log \left( \alpha_+ \, \alpha_- \,\left((u_0+v_0)\dfrac{\delta}{\delta+1} \right)^{\frac{\delta-1}{\delta}} \right)\,,
\end{equation}
and as we see, the dependence on $\alpha_{\pm}$ will cancel with that of the counterterms above. The entropy along the horizon in this case will be

\begin{equation}\label{scero}
S= -\dfrac{2\delta}{\delta+1} \dfrac{V_{{\bf R}^{d-1}}}{4G}(u+v_0)\, ,
\end{equation}
so we may rewrite again everything as a function of the entropy
\begin{eqnarray}
\label{kasnerYGH}
I_{\text{YGH}} &=& \dfrac{\delta+1}{2\pi \delta} (S_0/2-S)\, ,\\
2(I_{\Theta}^+ +I_{\Theta}^-)&=&\dfrac{S}{\pi \delta} \left[ 1+\log\left( \frac{\delta}{\delta+1}\, \frac{V_{{\bf R}^{d-1}}}{2G}\right) +\delta \log( \delta \,\LL )-\log S \right] - \dfrac{1}{2} \left( S \leftrightarrow S_0 \right)\, , \\
I_{\CV_u} &=& \dfrac{S}{\pi\delta} \dfrac{\delta-1}{2}\left[ \log\left(\dfrac{V_{{\bf R}^{d-1}}}{2G}\right) -\log S\right]\, ,\\
I_{\southcorner} &=& \dfrac{S_0}{ \pi\delta} \dfrac{\delta-1}{4}\left[ -\log\left(\dfrac{V_{{\bf R}^{d-1}}}{2G}\right) +\log S_0\right]\,,\label{kasnertip}
\end{eqnarray}
where $S_0$ is a constant that stands for the entropy evaluated at the south tip of the WdW patch $(u_0, v_0$). As we see, we recover the same structure for the counterterm as in the expanding case \eqref{TWct} and the black hole (see \eqref{BHct} in appendix \ref{section:BH})but in which the constant $\delta$ seems to play now the role of the codimension-2. Interestingly such effective dimension can in fact recover the value $\delta = d$ by considering the `holographic coordinate' $x_1$ to be the ripping direction in the most isotropic case, i.e. $p_1 = \frac{1-d}{d+1}$ and $p_{i \neq 1}=\frac{2}{d+1} $. Such values for the Kasner exponents correspond furthermore to the near-singularity approximation of the Schwarzschild black hole metric.

From equations \eqref{kasnerYGH} to \eqref{kasnertip}, we see that we obtain a log-linear locking for the total complexity density in the Kasner solution. Particularly, we can extract a growth rate of the form
\begin{eqnarray}\label{entlock}
\delta \widetilde{\CC}^* &=&  \left( a\, + b\log \tilde{S} \right) \delta \tilde{S}\, ,
\end{eqnarray}
with
\begin{eqnarray}
a &=&\dfrac{1}{\pi \delta} \left[ -\delta + \dfrac{\delta+1}{2} \log \left(\dfrac{1}{2G}\right) +\log\left( \dfrac{\delta}{\delta+1} \right)+ \delta\log(\delta \, \LL )  \right]\, ,\\
b &=& -\dfrac{\delta+1}{2 \pi \delta}\,  ,
\end{eqnarray}
where we have restored the  entropy density $\tilde{S}=S/V_{{\bf R}^{d-1}}$. 

Since $\delta >0$ we find that the coefficient of the logarithmic term, $b$, is always negative. Hence, the complexity density decreases sharply as ${\tilde S} \rightarrow 0$, in a log-enhanced example of entropic locking. On the other hand, the strict ${\tilde S} \rightarrow 0$ limit is likely to receive strong Planckian corrections and it is more natural to restrict the study of \eqref{entlock} to the region ${\tilde S} >1$. 

Under the weak assumption that the expansion length scale should not be transplanckian, i.e. $\ell_\Theta > \ell_{\rm Planck} \sim G^{1/d}$, the $a$ coefficient is dominated by the $\log (1/G)$ term and remains positive. This implies that \eqref{entlock} itself is dominated by the linear term,
\begin{equation}
\delta \widetilde{\CC}^* \approx a\,  \delta \tilde{S}\, ,
\end{equation}
within the interval $1<{\tilde S} \ll G^{-1}$. Using  \eqref{scero} we see that upper limit of this interval coincides with the condition that the Kasner patch be small in units of the Hubble rate, namely ${\tilde S}_0 \ll G^{-1}$. 

In conclusion, there is a standard linear entropic locking as ${\tilde S}$ decreases towards the Planckian regime ${\tilde S} \rightarrow 1$, and
the complexity decrease is further enhanced logarithmically if we insist on a (presumably unwarranted) extrapolation to subplanckian entropies. 
In any case, the phenomenology of entropic locking, namely that the monotonicity of the complexity is determined by that of the entropy, is found to hold in the local case as well.

\section{Conclusions}
\label{sec:conclusions}

In this paper we have studied exact solutions for which the limiting value of AC complexity depends on the entropy, already at the level of the leading bulk approximation. The need to consider situations with non-trivial behavior of the entropy requires paying careful attention to the expansion counterterm which ensures reparametrization invariance in the AC prescription. 

Our analysis can be seen as a complete definition of the terminal AC introduced in \cite{terminals}. In such an interpretation, the entropy is to be measured on the boundary of the causal domain of dependence of the singular set ${\cal S}^*$. We study exact solutions with divergent entropy in this sense, given by Coleman-de Luccia type solutions, and also patches of the  Kasner spacetime with vanishing entropy at the singularity, representing the local description of generic singularities in GR. 

We find the remarkable result that the terminal AC approaches a unified linear form in terms of the entropy,  in these two very different situations:
\begin{equation}
\delta \CC^*  \approx a \, \delta S  + \dots \,,
\end{equation}
up to $u$-independent constants and subleading terms. 
Although the detailed form of the coefficient $a$  depends on the particular solution, some general properties are to be noticed. In particular, it is interesting to notice that we find the general behavior
\begin{equation}
a = a_1 + a_2\,\log \LL\;,
\end{equation}
in appropriate units\footnote{ These are given by the AdS radius of curvature for the expanding bubble examples, and by the inverse Hubble scale $H^{-1}$ for the Kasner example.}.  Both $a_1$ and $a_2$ are strictly positive constants for all dynamical scenarios, therefore relegating negative values of $a$ only to those choices of $\LL$ that correspond to UV scales ($\LL \ll 1$) in the case of the expanding scenarios or even transplanckian ($\LL \ll \ell_{\text{Planck}}$) for the local case. Remarkably, this class of solutions comprises the most radical example of sensitivity of complexity to $\LL$ as they are the only known ones for which the late-time dynamics is qualitatively affected by the size of this scale. If one is to believe that the monotonic growth must be a generic property of terminal complexity in the expanding cases, the result above forces us to consider $\LL$ as an IR scale of the same order or lower than the  lowest scale present in the CFT, i.e. the curvature radius of the boundary metric.

Our results suggest that the program of classifying GR singularities according to their inherent complexity, an effort which goes back to \cite{Penrose, misner, BKL, BKL2, BKL3 } , acquires an interesting outlook when combined with holographic ideas: AC complexity seems to provide the required language. 
First, it was found  in \cite{terminals} that the  YGH term evaluated at the singularity defines a `complexity density' which serves as a holographic version of the Weyl curvature criterion by Penrose. In particular, we show that this contribution can be isolated  from the terminal AC complexity by a coarse-graining procedure, and we explicitly check that this quantity sets apart `simple' singularities, such as the one at a FRW bang, from `complex' ones, such as the generic black hole singularity.  

Second, we have seen that an  `entropic locking' of AC complexity arises when the entropy has strong dynamics near a spacelike singularity. On general grounds, we can imagine that the complexity grows linearly within a fixed Hilbert space, but it may have more complicated dynamics when the effective dimensionality of the Hilbert space, of order ${\rm exp}(S)$, changes abruptly with time. This was the situation found in \cite{rabinovc} in cases where the complexity was dominated by strong time-dependence of UV degrees of freedom. In the situations described in this paper, we are only concentrating in IR sectors, in the holographic sense, but again the effective Hilbert spaces have strongly time-dependent dimensionality and this phenomenon dominates the rate of change of complexity.

\section{Acknowledgments}
\label{ackn}

We would like to thank  C. Gomez, J. Magan, E. Rabinovici, M. Sasieta, R. Shir and  R. Sinha  for discussions on various aspects of computational complexity. This work is partially supported by the Spanish Research Agency (Agencia Estatal de Investigaci\'on) through the grants IFT Centro de Excelencia Severo Ochoa SEV-2016-0597,  FPA2015-65480-P and PGC2018-095976-B-C21. The work of J.M.G. is funded by 
Fundaci\'on La Caixa under ``La Caixa-Severo Ochoa'' international predoctoral grant. 


\appendix
\section{Terminal complexity of the eternal AdS black hole}
\label{section:BH}
\noindent

In this appendix we implement the reparametrization-invariant prescription for the quasilocal AC in the benchmark case of an eternal AdS black hole. We have included this discussion because the precise definition is not devoid of subtleties and requires a regularization procedure with careful account of orders of limits. 

The metric of a neutral and static AdS$_{d+2}$ black hole is 
\be\label{bhs}
\text{d}s^2 = -f(r)\,\text{d}t^2 + {\text{d}r^2 \over f(r)} + r^2 \,\text{d}\Omega_{d}^ 2\;, \qquad f(r) = k+ r^2 - {r_h^{d-1}(r_h^2+k) \over r^{d-1}}\;,
\ee
where $\text{d}\Omega_{d}$ stands for the spatial $d$-dimensional boundary metric and we measure length in units of the AdS curvature radius $\ell =1$. The constant $k$  takes the values $0,1,-1$ respectively for flat, spherical and hyperbolic boundary metrics. 

The total complexity of these solutions has been studied in detail including its early and late time dynamics and the influence of the action counterterms (cf. \cite{Myerstdep}). The direct application of \eqref{Ict} to the quasilocal terminal complexity must however be taken with care. As black hole horizons enjoy vanishing expansion along the horizon, the computation of the counterterms yields an identically null contribuition from the past WdW boundaries, preventing the cancellation of the null vector normalization constants. This seems to indicate that the definition of \eqref{Ict} does not hold properly for null expansion surfaces and might need to be replaced with a different expression. On the other hand, it can be checked that a regularization procedure can converge properly to a meaningfull result where the cancellation of ambiguities is guaranteed by taking the correct order of limits.

\begin{figure}[H]
\begin{center}
\includegraphics[width=7cm]{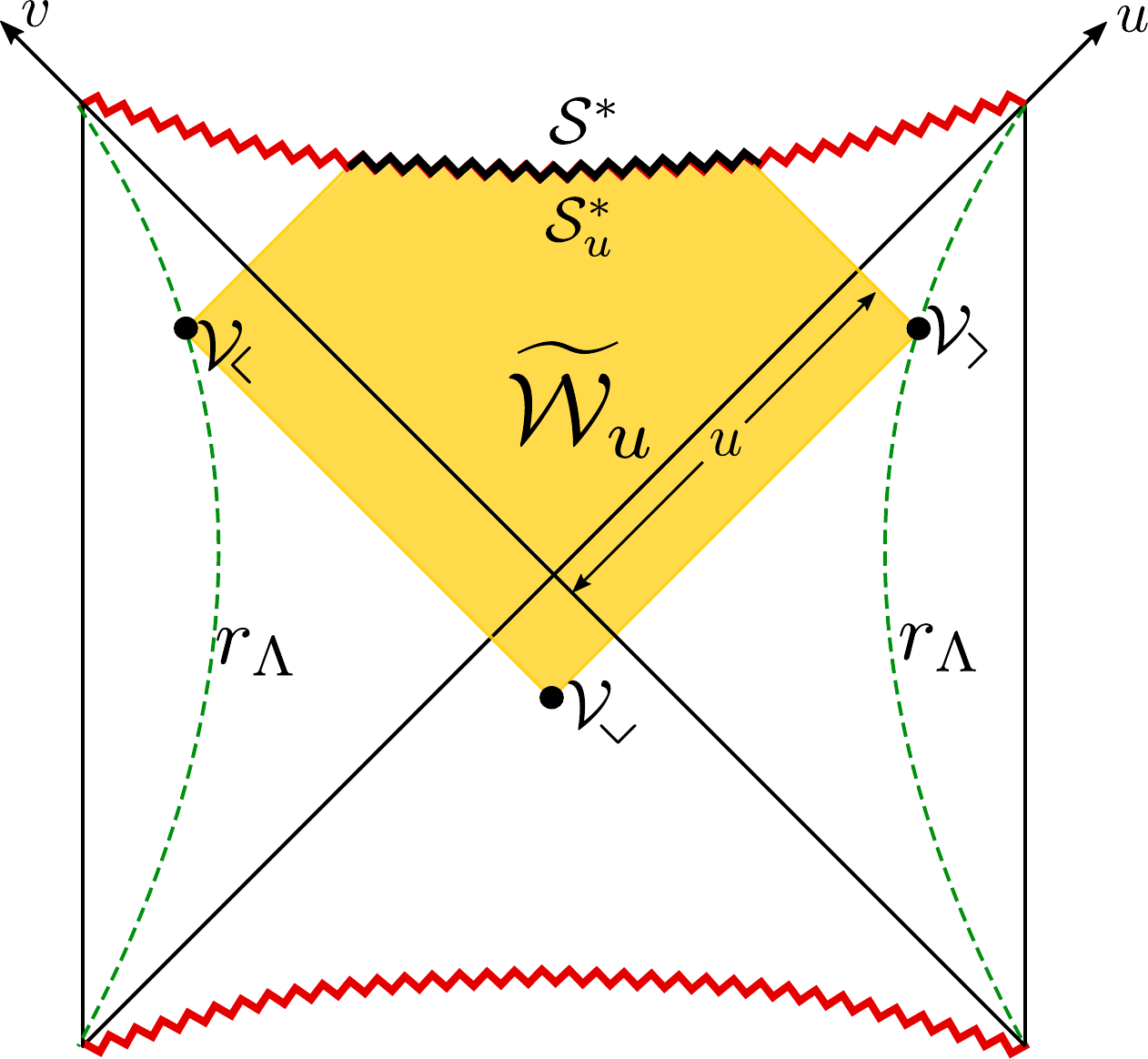} 
\caption{\emph{ Wheeler-deWitt patch $\widetilde{\CW_u}$ achored on the regularization surface $r_\Lambda$. The terminal complexity for the singularity slab ${\CS}^*_u$ is defined as the limit $\CC^*_u= \lim\limits_{r_\Lambda \rightarrow r_h} \widetilde{\CW_u}$ for fixed null coordinate $u$ along the horizon.}}
\label{fig:blackhole}
\end{center}
\end{figure}

Accordingly, we will take advantage of previous results and use them to argue that the terminal complexity for the black hole interior can be obtained as a proper limit from the complete AdS-Schwarzschild on-shell action. In particular, we may consider the terminal complexity at some nested WdW patch ${\CW}_u$  as the limit of the full complexity in which the regularization surface $r_\Lambda$ (see Figure \ref{fig:blackhole}) approaches the horizon $r_\Lambda \rightarrow r_h$ for a fixed value of the null coordinate $u$  along the horizon. A natural choice for such null coordinate is given by the compact Kruskal extension
\begin{eqnarray}
\tan u &=& \pm e^{2 \pi T(r_*(r)+t)}\,, \\ 
\tan v &=& \pm e^{2 \pi T(r_*(r)-t)}\,,  
\end{eqnarray}
where $r_*(r)$ is defined as usual as $dr_*=f(r)dr$ and $T$ is the Hawking temperature along the horizon. The signs must be taken in each quadrant in consistency with Figure \ref{fig:blackhole}. 

The full expression for the total complexity depends only on the Schwarzschild time $t$ and the position of the south tip $r_m$ which in turn we shall write in terms of $u$ and $r_\Lambda$. For any reflection-symmetric WdW patch $\CW_{u}$, it is easy to find the relation between the corners, given by
\begin{equation}
\log (\tan u) = {4 \pi T \left(r_*(r_\Lambda) - \frac{1}{2}r_*(r_m) \right)}\,,
\end{equation}
from which we may solve for $r_m(r_\Lambda, u)$. Unfortunately, the latter equation is in general not invertible, as the integral defining $r_*(r)$ is not always analytically solvable. However, as we will only care about the asymptotic late time regime, we can observe that $r_m \sim r_h$ and make use of the Rindler approximation
\begin{equation}
f(r) \simeq 4 \pi T (r-r_h) + \dots\,,
\end{equation}
so that we can approximate the integral to be
\begin{equation}
r_*(r) \simeq \int \dfrac{\text{d}r}{4 \pi T (r-r_h)} = \dfrac{1}{4 \pi T} \log \left(\dfrac{|r-r_h|}{\zeta r_h }\right) \,,
\end{equation}
with $\zeta$ a dimensionless integration constant. Inverting now this relation we get
\begin{equation}
r = r_h(1 - \zeta e^{4 \pi T r_*})\,,
\end{equation}
and we can find the position of the south tip as 
\begin{equation}
\label{rm}
r_m = r_h\left(1-\dfrac{1}{\tan^2 u} \dfrac{1}{\zeta r_h^2}(r_\Lambda-r_h)^2 \right).
\end{equation}

We are thus now ready to take the limit in order to obtain the terminal complexity for the black hole interior. In fact, we will check that such definition yields a finite and well behaved quantity, free of divergencies and ambiguities of any sort. Whereas the limit for the bulk and YGH contributions trivially leads to the finite quantities calculated in \cite{terminals}, we must check explicitly the joints and counterterm pieces of the action. The evaluation of the three  non-trivial joints in the WdW patch leads
\begin{eqnarray}
\label{southjoint}
I_{\CV_{\southcorner}} &=& -\dfrac{V_\Omega r_m^{d}}{8\pi G } \log|f(r_m)|\,, \\ 
I_{\CV_{\eastcorner}}= I_{\CV_{\westcorner}} &=& \dfrac{V_\Omega r_\Lambda^{d}}{8\pi G } \log|f(r_\Lambda)|\,,
\label{eastwestjoints}
\end{eqnarray}
where we omitted here the dependence on the null vectors normalization constants, which has been shown to cancel after all counterterms are added. As we see, the three contributions are separately divergent as $r_\Lambda \rightarrow r_h$. Feeding however \eqref{rm} into \eqref{southjoint} and approximating also $f(r_\Lambda) \simeq 4 \pi T (r_\Lambda-r_h)$ in \eqref{eastwestjoints} we see that these divergences cancel out, yielding a total contribution
\begin{equation}
\label{Jointspiece}
I_{\CV_{\southcorner}}+I_{\CV_{\eastcorner}}+I_{\CV_{\westcorner}}\simeq \dfrac{S}{2 \pi} \log \left(4 \pi T \zeta r_h \tan^2 u  \right),
\end{equation}
which is finite and contributes to the complexity growth as usual (cf. \cite{Fischler}) with a piece proportional to the black hole entropy

\begin{equation}
S= \frac{V_\Omega r_h^d}{4G}\,.
\end{equation}

We can also calculate the contribution from the four null boundary counterterms, which gives (cf. \cite{Myerstdep})
\begin{equation}
I_{\Theta}= \dfrac{V_{\Omega}}{2\pi G d}r_\Lambda^{d} \left(1+ d\log \left( \dfrac{d \, \LL}{r_\Lambda} \right) \right) -\dfrac{V_{\Omega}}{4\pi G d}r_m^{d}\left(1+ d\log \left(\dfrac{d\,\LL}{r_m} \right) \right).
\end{equation}
The limit of this quantity as $r_\Lambda \rightarrow r_h$ gives us trivially
\begin{equation}
I_{\Theta}= \dfrac{S}{\pi d} \left(1+d \log \left( \dfrac{d\, \LL}{r_h} \right) \right) ,
\end{equation}
or, purely as a function of the entropy
\begin{equation}
\label{BHct}
I_{\Theta}= \dfrac{S}{\pi d} \left(1+\log\left(\dfrac{V_\Omega}{4G}\right)+ d\log \left( d \,\LL \right) -\log S \right) ,
\end{equation}
which is of course constant and therefore does not contribute to the complexity growth rate. Taking now the bulk and YGH contributions to terminal complexity (cf. \cite{terminals}) we can write finally the full rate 
\begin{equation}
\label{BHrate}
\dfrac{d \CC^*}{du} =  \dfrac{1}{\cos u \sin u} \;\dfrac{2M}{2 \pi T}\,,
\end{equation}
and the standard result in terms of the Schwarzschild time, $d \CC^* /dt=2M $, can be readily  recovered using the chain rule. In this case however the result holds not only for late times but exactly for all times provided the WdW patch touches the singularity\footnote{As the action of the south corner appears with a logarithm, its contribution to the dynamics survives the limiting procedure, a fact that one may consider spurious from the direct definition of the domain $D^-(\mathcal{S}^*)$. Removing this contribution amounts to a shift $2M \rightarrow 2M-TS$ in \eqref{BHrate} but does not modify the qualitative behavior of complexity growth, which remains positive anyway.  }. As $u \in (0,\pi/2)$ this rate is always positive, ensuring the monotonic growth of holographic complexity. For asymptotically late times $u \sim u_*=\pi/2$ we get

\begin{equation}
\dfrac{d \CC^*}{du} \simeq  \dfrac{1}{(u_*-u)} \, \dfrac{2M}{2 \pi T}\,.
\end{equation}
\newpage


\bibliographystyle{utphys.bst}
\bibliography{refs}{}

%
%

\end{document}